# New Estimates of the Solar-Neighborhood Massive Star Birthrate and the Galactic Supernova Rate

B. Cameron Reed

Department of Physics
Alma College
Alma, MI 48801

ph: (989) 463-7266
fax: (989) 463-7076
e-mail: reed@alma.edu

# ABSTRACT

The birthrate of stars of masses $\geq 10$ M$_\odot$ is estimated from a sample of just over 400 O3-B2 dwarfs within 1.5 kpc of the Sun and the result extrapolated to estimate the galactic supernova rate contributed by such stars. The solar-neighborhood galactic-plane massive star birthrate is estimated at $\sim 176$ stars/kpc$^3$/Myr. Based on a model where the galactic stellar density distribution comprises a "disk + central hole" like that of the dust infrared emission (Drimmel & Spergel 2001) the galactic supernova rate is estimated as probably not less than $\sim 1$ nor more than $\sim 2$ cy$^{-1}$ and the number of O3-B2 dwarfs within the solar circle as $\sim 200,000$.

**Subject Headings:**   stars: early-type --- supernovae: general

1. **Introduction**

All main-sequence stars of 10 $M_\odot$ or more (spectral types earlier than ~B2) are believed to end their lives as supernovae (SNe) by one evolutionary process or another (Massey 2003). As supernovae are significant contributors to the structure, kinematics, and enrichment of the interstellar medium, to the flux of cosmic rays, and to the birthrates of pulsars, runaway stars, and stellar black holes, such massive stars have astrophysical consequences far out of proportion to their small numbers relative to the much more common lower-mass stars.

Because the number of historically-recorded supernovae in the Galaxy lies in the single digits, determining their birthrate has always been a highly uncertain endeavor. Most estimates, however, place that number at a few per century ($cy^{-1}$). Based on surveys of radio remnants and known supernovae within the last 1000 years, van den Bergh & Tammann (1991; vdB&T) estimated rates of 3.4 $\pm$ 2.0 $cy^{-1}$ and 5.8 $\pm$ 2.4 $cy^{-1}$, respectively, and remarked that the current mass spectrum of star formation would lead one to infer a rate of ~ 2 $cy^{-1}$. Cappellaro et. al. (1993) [see also Wheeler & Benetti (2000)] derived a rate of 1.7 $\pm$ 0.9 $cy^{-1}$ based on applying *external galaxy* rates to the Milky Way and assuming a galactic blue luminosity of (2.0 $\pm$ 0.6) x $10^{10}$ $L_\odot$. However, as vdB&T discuss, determination of extragalactic supernovae rates is fraught with a variety of selection effects.

The purpose of this paper is to provide an independent estimate of the galactic supernova rate based on a direct census of massive (M $\geq$ 10 $M_\odot$) early-type main-sequence stars within the solar neighborhood combined with assumed models for extinction, the spatial distribution of these stars, and their main-sequence lifetimes. While this approach is not without its own limitations, it does have the advantage of deriving from what is believed to be a quite complete inventory of such objects to the magnitude limit adopted and so should complement the approaches summarized above. The present results can be expected to capture those SNe that arise from "core-collapse" (those of Types Ib, Ic, and II) but not those arising from deflagrating CO white dwarfs that are thought to lead to Type Ia SNe. More detailed descriptions of the various SNe Types can be found in vdB&T and Wheeler and Benetti (2000).

The approach taken here is similar to that used to derive the luminosity function (LF) of solar-neighborhood OB stars in an earlier paper (Reed 2001; hereafter Paper I). Since the absolute magnitudes of early-type dwarfs overlap with those of their more evolved counterparts, one cannot simply extract some limited magnitude range of the LF from which to estimate the stellar birthrate. Rather, one must begin by isolating a magnitude-limited sample of main-sequence stars down to temperature class B2 that have reliable photometry and spectral classifications available. The procedure is then to (i) adopt a model for the interstellar absorption in order to transform the limiting apparent magnitude into limits on distance as a function of absolute magnitude, (ii) to adopt a model for the local spatial distribution of these stars to facilitate computing a total *effective* number of stars out to the distance of the most luminous members of the sample, and, finally, (iii) by adopting a model for stellar main-sequence lifetime as a function of absolute magnitude, transform the effective number of stars into an average stellar birthrate in units of stars per kpc$^3$ per million years. This result can be roughly extrapolated to the entire galaxy to produce a supernova-rate estimate.

This approach is subject to a variety of uncertainties, each difficult to quantify in its own right. Five main sources of uncertainty can be identified: (1) Some bright OB stars still lack spectral classifications. However, for the magnitude limit adopted in the present work (V = 8), this amounts a small number of stars (~ 30 out of nearly 2700) and the number of these that might be dwarfs can be reasonably estimated by extrapolating from the percentage of classified stars that are thought to be so. (2) While distances for individual stars can be computed on the basis of spectroscopic parallaxes, it is necessary to assume a simple global extinction model in order to establish limiting distances as a function of limiting apparent and absolute magnitudes; such a simple extinction model cannot hope to capture the erratic nature of the interstellar absorption around the sky. (3) It is necessary to assume an *a priori* model for the local density distribution of these stars in order to normalize their counts as a function of absolute magnitude to a uniform limiting distance; this is because these stars are not uniformly distributed in space and the distances involved correspond to several times the scale height of their distribution perpendicular to the galactic plane. While the model adopted is well-substantiated, it does represent a condition imposed upon the data. (4) Similarly, it is necessary to assume a model for the main-sequence lifetime of stars as a function of their absolute magnitudes. Even if one were to have complete knowledge of the stellar population to some limiting distance, that population will include objects whose main-sequence lifetimes vary by a factor of over six (see Table 1), and it is necessary to normalize to some timescale in order to establish a meaningful presumed-steady-state birthrate. (5) Finally, there is an underlying confounding astrophysical issue: as discussed by Massey (2003), it is misleading to think that there exists a unique mass-spectral type relationship for the most massive stars, but that one should rather think of spectral types as stages through which stars of different mass pass in their evolution. In an example given by Massey, a star of spectral type O4 V may be either a zero-age 60 $M_\odot$ star or a slightly older 85 $M_\odot$ star. As there is no way to independently determine a star's mass or age from the observational material used in this work there is no unique way to resolve this issue; for purposes of assigning main-sequence lifetimes I make the blanket assumption that all stars are zero-age main-sequence objects.

Other complications lurk beyond these issues: as Massey points out in his Sections 1 and 2, the evolution of massive stars is difficult to model owing to complications such as mass loss, rotation, convection, and metallicity. Also, it should be noted that Hansen et al. (2004) give the lower mass limit for stars that become core-collapse SNe as 8 $M_\odot$, about that of a B3 star. However, as the OB-star catalog on which this work is based becomes seriously incomplete beyond type B2, I adopt a ~10 $M_\odot$ cutoff.

The sample is described in section 2 below along with the assumed reddening, magnitude-lifetime, and spatial distribution models. The resulting stellar birthrate and extrapolated galactic supernova rates are discussed in section 3. A brief summary appears in section 4.

## 2. Data and Model Assumptions

### 2.1 OB-star Catalog and Data

This author maintains a catalog of and databases tabulating published UBVβ photometry and MK spectral classifications for galactic OB stars. These are described in Reed (2003) and references therein; that paper also gives a brief history of their development. Instructions on accessing the catalog, files, and supporting documentation are available at http://othello.alma.edu/~reed/OBfiles.doc. Based originally on objects cataloged in the Case-Hamburg galactic-plane surveys (to galactic latitudes $\pm 10°$), this work is now a fully all-sky effort. 'OB stars' here is taken to mean main-sequence stars down to temperature class B2 and more luminous ones down to temperature class B9. The original Case-Hamburg surveys (about 12,000 stars) also include some 2000 evolved A-G stars along with some white dwarfs, planetary nebulae, and Wolf-Rayet stars. It is worth noting that the definition of an OB star is not universal; for example, Vanbeveren et al. (1998) define them as O-B2 V-IV, O-B3 III, O-B4 II and all OBA Ib, Iab, Ia, and IaO stars. These differences in definition are not significant in the present context, however, as we are concerned only with main-sequence stars.

At this writing (June 2005), my catalog lists nearly 18,400 independent objects for which some 38,000 photometric observations and 22,900 spectral classifications are available. Several thousand of the fainter stars yet lack fundamental data. Only a small fraction of these 18,000-odd stars are ultimately utilized in this work.

Two completeness issues attend the present work. The first is that full MK classifications (both temperature and luminosity types) are not available for all stars for which photometry is available. For example, UBV photometry is available for 7281 stars to V = 10, but only 6314 of these (87%) have MK classifications available. This issue can be dealt with by estimating the number of unclassified stars that could be expected both to be main-sequence stars and lie within the distance limits adopted by extrapolating from the number of fully-classified stars that are and do so (see section 2.6). The second issue is more difficult: How complete are my catalogs to a given limiting apparent magnitude? This is answered *a posteriori*. As described below (section 2.5), one of the results determined in this work is the present-day galactic-plane density of O3-B2 dwarfs, $\rho_{plane}$. If the catalogs are complete, $\rho_{plane}$ should remain fairly constant as fainter limiting magnitudes are investigated. The limit of completeness is consequently taken to be where $\rho_{plane}$ begins to drop, which proves to be just beyond V ~ 8 (see Figure 2).

## 2.2     Spectral Classifications and Sample Selection

The approximately 23,000 spectral classifications in my database are drawn from over 400 sources. As might be expected, these classifications derive from photographic and electronic material obtained with a variety of instruments employing various dispersions and wavelength coverage. Since the present work depends on deriving stellar distances via spectroscopic parallaxes, isolating the best-quality classifications is essential. Consequently, all sources were examined and assigned a 'classification quality code' from A to E, with A being the highest. In brief, code A designates classifications that adhere to the original MK criteria of wavelength coverage ~ 3500-5000Å, dispersion 60-125 Å/mm, and resolution 1-2Å. Code B designates classifications that nominally adhere to these criteria but which are suspected of possibly being of slightly lower quality, such as objective-prism spectra where obtaining the optimum exposure for each star can be difficult. Code C designates classifications derived from material that was obtained for classification purposes but which failed to meet the MK criteria in some way; in many cases these were high-dispersion objective-prism or 'thin prism' classifications. Code D designates classifications deriving from material failing to meet the MK criteria and which was obtained for some other purpose such as radial velocity or abundance studies, and code E is a catch-all category reserved for cases where authors give little or no detail on their instrumental system and so whose classifications could not be meaningfully coded.

My original intent had been to use only classifications of quality A, B, or C with some arbitrary weighting scheme. However, as described in section 3 below, the vast majority of classifications adopted proved to be of type A, so no further discrimination was felt necessary.

A program was written to merge my photometric and classifications files and to output a list of highest-quality useable classifications for all stars brighter than some operator-assigned limiting apparent magnitude $V_{lim}$. 'Useable' here means a classification with both full temperature and luminosity components; in a few cases, lower-quality classifications were selected where nominally higher-quality ones lacked luminosity classes. If more than one highest-quality classification was found for a given star, each was treated separately. A second program was written to read in the output of the first program and assign absolute visual magnitudes (taken here to be synonymous with absolute magnitudes) and intrinsic B–V colors for the selected classifications from the calibrations of Turner (1980) and Schmidt-Kaler (1982), respectively, interpolating linearly across either or both of temperature and luminosity class where necessary. These magnitude and color calibrations are the same ones as were adopted in Paper I. A distance corresponding to each classification is computed via spectroscopic parallax, and the program outputs a list of magnitudes, classifications, and distances for O3-B2 main-sequence stars. While Walborn, et al. (2002) have defined criteria for classifying O2 stars, only a handful of such stars appear in the my databases; as Turner (1980) does not give absolute magnitudes for this type I decided not to include them in this work in order to maintain consistency with paper I. In any event, there are no O2 or O3 stars brighter than the V = 8 limiting magnitude I ultimately adopted.

## 2.3 Reddening Model and Distance Limits

All of the absolute magnitudes assigned by the second program described above fall into one of about 20 discrete values, reflective of the fact that temperature classes are usually assigned with no more precision than 0.5 subclasses. The next step is, for each of these $M_V$, to transform $V_{lim}$ into a corresponding limiting distance. Since even at $V_{lim}$ = 11 only some 1900 OB dwarfs emerge from the second program, the number of stars per square degree of sky is too small to attempt any meaningful extinction analysis in any sort of sector-by-sector-of-sky scheme. I therefore adopt, as in Paper I, a global absorption model of extinction 1.5 mag/kpc. For each $V_{lim}$, distances corresponding to the intrinsically brightest (O3; $M_V$ = –5.4) and faintest (B2; $M_V$ = –2.2) sample members were computed; for computational purposes, the limiting distance for any $M_V$ is then taken to be given by the linear relationship

$$\log (r_{max}^{M_V}) = A + B(M_V) \qquad (1)$$

where A and B are derived by fitting equation (1) to the absolute magnitudes and distance limits of the O3 and B2 stars; A and B will be functions of $V_{lim}$. A third program was written to reduce the output of the second program to a list of OB dwarfs within the distance limits prescribed equation (1). In actuality, an extinction of 1.5 mag/kpc is a reasonable choice: In the case of $V_{lim}$ = 8 (the adopted limiting magnitude), the average extinction derived from 656 classifications for 421 stars within the distance limits of equation (1) is 2.1 ± 2.3 (s.d.) mag/kpc with a median value of 1.45 mag/kpc.

A side question that arises in the selection of stars and distance limits is whether or not a significant Malmquist bias might come into play. This effect is a tendency for stars (of a given type) in a sample selected by apparent magnitude to be intrinsically brighter than the average for all such stars in a given volume. The resulting difference in absolute magnitude can be shown to be $\Delta M \sim -2.3\sigma^2\{d[\log A(m)]/dm\}$ where $A(m)$ is the number of stars in apparent magnitude range $m \pm dm/2$ and $\sigma$ is the dispersion of the luminosity function for the type of stars involved (Mihalas & Binney 1981). I have examined this issue for 622 B2 dwarfs to V = 11, finding $d[\log A(m)]/dm \sim 0.09$; taking $\sigma \sim 0.5$ gives $\Delta M \sim -0.05$, a completely negligible effect in comparison with the other approximations and assumptions involved in this work.

## 2.4  Main Sequence Lifetime Model

To transform this inventory of OB dwarfs into a stellar birthrate it is necessary to adopt a relationship that gives a star's main-sequence lifetime ($\tau_{MS}$) as a function of its absolute magnitude.

The Geneva Observatory group has published an extensive array of stellar models over the past several years. For the present purposes, those given by Schaller et al. (1992) for solar metallicity ($Z = 0.02$) are most appropriate. For a range of masses that incorporates those of O3-B2 dwarfs, Table 1 lists main-sequence lifetimes, luminosities, and effective temperatures drawn from Tables 2-9 of Schaller et al. "Main-sequence lifetime" is taken here to be the duration of core hydrogen burning, that is, the age listed at "point 13" in the Schaller et. al. tables. The luminosities and effective temperatures are taken to be those at the start of the hydrogen-burning phase, "point 1" in their tables. The bolometric corrections listed in the fifth column of Table 1 were computed from the Vacca et. al. (1996) calibration

$$BC = 27.66 - 6.84 \log(T_{eff}), \qquad (2)$$

which they found to be an excellent fit for O3 to B0.5 stars; I assume that it can be extrapolated to B2 stars. The absolute visual magnitudes appearing in the last column of the table were computed assuming a solar bolometric magnitude of +4.72.

Figure 2 shows the values of Table 1 as a plot of $\log(\tau_{MS})$ vs. the absolute value of $M_V$. An unweighted least-squares power-law fit gives

$$\log(\tau_{MS}) \text{ (yr)} = 8.1607 \, |M_V|^{-0.1341}, \qquad (3)$$

with correlation coefficient $r^2 = 0.9994$. This relationship is used to assign a main-sequence lifetime for each star and hence to normalize the inferred birthrates of stars to the lifetime of the faintest stars in my sample.

## 2.5 Birthrate Calculation

The first step in calculating the local massive-star birthrate is to account for the fact that stars of different absolute magnitudes are sampled to different distance limits as given by equation (1). Let $N(M_V)$ be the number of stars of absolute magnitude $M_V$ within the distance limit $r_{max}^{M_V}$ given by equation (1), and let $r_b$ be the cutoff distance of the intrinsically *brightest* sample members as computed from equation (1). It is desired to normalize the star counts to distance $r_b$. However, contrary to what one might think, this is *not* given by scaling $N(M_V)$ by $(r_b/r_{max}^{M_V})^3$. This is because OB stars are not uniformly distributed in space. Rather, they are distributed (locally) about the galactic plane with an exponential density profile of the form

$$\rho = \rho_{plane} \exp(-|z|/h), \tag{4}$$

where $\rho_{plane}$ is the (present-day) solar neighborhood OB dwarf density in the galactic plane and h is a scale height. In an earlier paper (Reed 2000) it was found that $h \sim 45$ pc for OB stars. With this form of density profile, the predicted number of stars N over the entire sky out to limiting distance $r_{max}$ is given by [see equations (4) and (5) of Reed (2000) with $b = 90°$]

$$N = 2\pi \rho_{plane} \left\{ r_{max}^2 h(1-e^{-x}) - 2h^3[1-e^{-x}(x^2/2 + x + 1)] \right\}, \tag{5}$$

where

$$x = r_{max}/h. \tag{6}$$

In the present work the distance cutoff for the intrinsically faintest stars of my sample never falls below 344 pc (B2 stars; $V_{lim} = 6$), or about 7.6 scale-heights. This renders $e^{-x} \sim 0.0005$, hence, to an excellent approximation, equations (5) and (6) can be simplified to

$$N = 2\pi \rho_{plane} r_{max}^2 h. \tag{7}$$

Consequently, when scaled to cutoff distance $r_b$, the present-day number of stars of absolute magnitude $M_V$ is proportional to the *square* of the distance ratio:

$$N(M_V, r_b) = N(M_V) \left[\frac{r_b}{r_{max}^{M_V}}\right]^2. \tag{8}$$

Now let $\tau_{M_V}$ and $\tau_f$ be the main-sequence lifetimes of stars of absolute magnitude $M_V$ and of the intrinsically *faintest* stars in the sample, respectively. Assuming that the stellar population is in a steady state over time $\tau_f$, the *effective* number of stars of absolute magnitude $M_V$ over time $\tau_f$ out to distance $r_b$ will be

$$N_{eff}(M_V, r_b, \tau_f) = N(M_V) \left[\frac{r_b}{r_{max}^{M_V}}\right]^2 \left[\frac{\tau_f}{\tau_{M_V}}\right]. \tag{9}$$

The average solar-neighborhood massive-star birthrate is determined by summing the $N_{eff}$ over all absolute magnitude values, dividing by the volume corresponding to the cutoff distance $r_b$, and dividing by the timescale $\tau_f$:

$$\text{birthrate} = \frac{3}{4\pi r_b^3 \tau_f} \sum_{M_V} N_{eff}(M_V, r_b, \tau_f) = \frac{3}{4\pi r_b} \sum_{M_V} \frac{N(M_V)}{\tau_{M_V} \left[r_{max}^{M_V}\right]^2}. \tag{10}$$

More meaningful, however, is the birthrate in the galactic plane. From equation (7), the in-plane density $\rho_{plane}$ and average stellar density $<\rho>$ must be related as ($N = 4\pi r^3 <\rho>/3$)

$$\rho_{plane} = \frac{2}{3}\left(\frac{r}{h}\right)<\rho>, \tag{11}$$

hence the in-plane birthrate will be (2r/3h) times that of the average birthrate of equation (10). The (present-day) in-plane OB dwarf density is given by equation (7) with N as the present-day number of stars (equation 8; $r_b = r_{max}$)

$$\rho_{plane} = \frac{1}{2\pi r_{max}^2 h} \sum_{M_V} N(M_V, r_b) = \frac{1}{2\pi h} \sum_{M_V} \frac{N(M_V)}{[r_{max}^{M_V}]^2}. \qquad (12)$$

To extrapolate the local birthrate to the entire galaxy it is necessary to adopt a model for the run of stellar density distribution within the galactic disk. I adopt a two-component model: (i) a disk with both radial and vertical density gradients whose density increases toward the galactic center but which at some galactocentric radius $R_{hole}$ is superseded by (ii) a decreasing density of Gaussian form in the galactocentric radial coordinate R for $0 \leq R \leq R_{hole}$. The specific form adopted is

$$\rho(R,z) = \begin{cases} \rho_{disk}\, e^{-R/H} e^{-|z|/h} & (R > R_{hole}) \\ \rho_{hole}\, e^{-[(R-R_{hole})/R_{hole}]^2} e^{-|z|/h} & (R \leq R_{hole}) \end{cases} \qquad (13)$$

where H is the disk radial density scale length, h is the vertical scale height of the OB stars as in equations (4) et. seq. and $R_{hole}$ is the galactocentric radius at which the exponential disk begins. (R is measured parallel to the galactic disk from a z-axis through the galactic center.)

This "disk + hole" model is motivated by the work of Drimmel & Spergel (2001), who studied the distribution of dust in the Galaxy at infrared wavelengths. They found the dust emission to be well-represented by such a model with $H = 0.28\, R_o$ and $R_{hole} \sim R_o/2$ where $R_o$ is the Sun's distance from the galactic center (see their section 3.2 and Figure 13); it would seem plausible to assume that the massive star distribution should follow that of the dust. Drimmel & Spergel actually used a squared-hyperbolic-secant for the vertical part of their disk density distribution in addition to a global warp and spiral-arm segments; a purely exponential vertical distribution is adopted here for simplicity and to maintain consistency with Reed (2000). Of course, any one of an almost infinite number of physically plausible density distributions could be adopted; however, given the approximations and assumptions already built into the present work there seems little point in attempting any more sophisticated analysis.

Putting $z = 0$ and $R = R_o$, parameter $\rho_{disk}$ of equation (13) and the solar-neighborhood galactic-plane OB-star density $\rho_{plane}$ (equation 4) are related as

$$\rho_{disk} = \rho_{plane} \exp(R_o/H). \tag{14}$$

Demanding that the stellar density be continuous for all values of z at $R = R_{hole}$ gives (equations 13 and 14)

$$\rho_{disk} e^{-R_{hole}/H} = \rho_{hole} \quad \Rightarrow \quad \rho_{hole} = \rho_{plane} e^{(R_o - R_{hole})/H}. \tag{15}$$

The total number of OB stars within the solar circle at the present time is given by

$$N_{galaxy}(R_o) = 4\pi \int_{z=0}^{\infty} \int_{R=0}^{R_o} \rho(R,z) R\, dR\, dz$$

$$= 4\pi \rho_{plane} \left\{ e^{R_o/H}(hH^2)\left[ e^{-R_{hole}/H}(R_{hole}/H + 1) - e^{-R_o/H}(R_o/H + 1) \right] \right.$$

$$\left. + (hR_{hole}^2) e^{(R_o - R_{hole})/H} \left[ (\sqrt{\pi}/2)\, \text{erf}(1) - (1 - e^{-1})/2 \right] \right\} \tag{16}$$

where the first and second terms arise from the disk and hole contributions, respectively, and where erf denotes an error function. The limit on z here is formal: any choice of an upper limit on z that is several times the vertical scale height h will give essentially the same result. The time-derivative of this result can be used to estimate the galactic massive-star birthrate from the solar-neighborhood (galactic plane) birthrate.

## 2.6  Correcting for Classification Incompleteness

In this section I describe briefly how the calculations described above can be approximately corrected to account for stars lacking full MK classifications.

For limiting apparent magnitude $V_{lim}$, let N be the number of cataloged O3-B2 stars to apparent magnitude $V_{lim}$, $N_{MK}$ be the number of these that have full classifications, $N_{O3\text{-}B2}$ be the number that are O3-B2 dwarfs, and $N_D$ be the number within the distance cutoff of equation (1).

The first step is to account for those $N-N_{MK}$ stars without classifications. The fraction of these that could be expected to be O3-B2 dwarfs is estimated as $N-N_{MK}$ times the fraction of the $N_{MK}$ stars so classified: $(N-N_{MK})(N_{O3\text{-}B2}/N_{MK})$. The fraction of these "expected additional" O3-B2 dwarfs that are within the distance cutoff is then taken to be the ratio of the number of stars known to be within the distance cutoff to the total number known to be O3-B2 dwarfs, $(N_D/N_{O3\text{-}B2})$. Hence, we could expect to have $N_{extra} = (N-N_{MK})(N_{O3\text{-}B2}/N_{MK})(N_D/N_{O3\text{-}B2}) = (N-N_{MK})(N_D/N_{MK})$ additional stars within the distance cutoffs if all stars to magnitude $V_{lim}$ had full classifications available. These additional stars need to be propagated into the present-day plane-density and birthrate calculations. The density calculation is predicated on the projected number of present-day stars P within the distance limit of the intrinsically brightest stars, that is, $P = \Sigma\, N(M_V, r_b)$ where the sum runs over $M_V$; see equation (8). The number of extra present-day stars can then be expected to be

$$P_{extra} = N_{extra}\left(\frac{P}{N_D}\right) = (N - N_{MK})\left(\frac{N_D}{N_{MK}}\right)\left(\frac{P}{N_D}\right) = (N - N_{MK})\left(\frac{P}{N_{MK}}\right). \quad (17)$$

The correction factor CF is then defined to be

$$CF = \frac{P + P_{extra}}{P}. \quad (18)$$

Under the assumption that the local stellar population remains in a steady state, this correction factor is applied multiplicatively to both the derived density and birthrate results.

## 3. Results

A fourth program was written to perform the density and birthrate calculations described above. The sequence of four programs was then run for $V_{lim}$ = 6.0 (0.5) 11.0. For stars with multiple classifications, each was weighted equally as 1/n where n is the number of classifications involved for that star. Because of this, it is possible to have fractional real and effective numbers of stars at a given $M_V$.

Figure 2 shows the resulting corrected solar-neighborhood galactic-plane OB star density as a function of limiting magnitude. The error bars are based on Poisson errors corresponding to the number of O3-B2 dwarfs within the distance limits of equation (1) for each limiting magnitude. Clearly, the derived density begins to drop noticeably beyond $V_{lim}$ = 8, which I adopted as the limiting magnitude. Detailed *uncorrected* statistics for $V_{lim}$ = 8 are given in Table 2. Of 2666 stars in my databases with V ≤ 8, a total of 2584 have classifications available. 520 of these proved to be O3-B2 dwarfs, of which 421 emerged as lying within the adopted distance limits; a total of 656 classifications are available for these 421 stars. Of these 656 classifications, 616 are of quality code A and 19 of code B; consequently, it was felt that no code weighting was necessary. A file of star identifications, B and V values, classifications and derived distances can be downloaded from http://othello.alma.edu/~reed/N=656.dat.

As indicated in Table 2, I estimate that an uncorrected total of 1577 O3-B2 stars presently lie within the 1510 pc cutoff distance of the O3 stars; from the numbers given in the preceding paragraph and equations (17) and (18) this leads to a correction factor of 1.0317 for a corrected number of 1627 present-day stars. This yields a corrected average density of ~ 113 stars/kpc$^3$ and a corrected galactic-plane density of 2524 stars/kpc$^3$. Over the lifetime of a B2 star a corrected effective total of 2492 stars are involved, which yields a mean birthrate of 7.86 stars/kpc$^3$/Myr and a galactic-plane birthrate of ~ 176 stars/kpc$^3$/Myr.

Adopting $R_o$ = 8.5 kpc, H = 0.28$R_o$, h = 45 pc and $R_{hole}$ = 4.25 kpc, equation (16) gives $N_{galaxy}(R_o)$ ~ 163,600; the total birthrate evaluates to ~ 1.14 cy$^{-1}$. These numbers can be increased (or decreased) by decreasing (or increasing) $R_{hole}$; for example, $R_{hole}$ = 2 kpc gives $N_{galaxy}(R_o)$ ~ 229,000 and a birthrate of ~ 1.6 cy$^{-1}$ (again assuming $R_o$ = 8.5 kpc, H = 0.28$R_o$ and h = 45 pc). In the extreme case that $R_{hole} \to 0$, I find $N_{galaxy}(R_o)$ ~ 250,600 and a birthrate of ~ 1.75 cy$^{-1}$. That this latter result corresponds closely with the estimates of Cappellaro et al. (1993; 1.7 cy$^{-1}$) and vdB&T (1991; ~ 2 cy$^{-1}$ based on the current mass spectrum of star formation) may suggest that $R_{hole}$ for the OB stars is not as large as $R_o/2$. However, given the numerous assumptions built into the present work one must be careful to not overinterpret such a comparison. Also, the present results would incorporate only the "core-collapse" SNe of Types Ib, Ic, and II as opposed to the deflagrating CO white dwarfs thought to lead to Type Ia SNe. According to figures given by Cappellaro et al. (1993), some 80% of galactic SNe are anticipated to be of Types Ib/c and II. Any reasonable choice for $R_{hole}$ will lead to a SNe rate of ~ 1–2 cy$^{-1}$.

One clear conclusion is that many more galactic OB stars probably await discovery. The number of such stars estimated here is considerably larger than that derived in Paper I (~18,600) due to the present work deriving from an all-sky survey incorporating some 6,000 more stars than were used in that paper. Since the bulk of the additional stars are early-B dwarfs, their incorporation would not much alter the LF derived in Paper I for the intrinsically brighter OB stars.

4.  **Summary**

New estimates of the local massive-star birthrate and galactic supernova rate have been derived from a sample of ~ 420 O3-B2 dwarfs within about 1.5 kpc of the Sun to a limiting magnitude V = 8. Depending on values adopted for parameters describing the galactic stellar density distribution, the galactic supernova rate evaluates as 1–2 $cy^{-1}$; a result in line with other estimates.

Further progress in quantifying the galactic massive-star birthrate by this sort of direct star-counting is largely a matter of both continuing to search for previously undiscovered galactic OB stars and of obtaining photometry and spectral classifications for those already cataloged but which have not been followed up: nearly 1000 OB stars with V $\leq$ 10 still lack spectral classifications. In this regard, efforts such as forthcoming northern-hemisphere volumes of Nancy Houk's Michigan Spectral Survey should prove invaluable.

I am grateful to Tony Moffat, whose comments on this paper led to a number of significant improvements.

**Figure Captions**

Figure 1

Main-sequence lifetime vs. absolute value of absolute visual magnitude; see Table 1. The solid line shows the adopted relationship, equation (3).

Figure 2

Present-day solar-neighborhood galactic-plane O3-B2 dwarf density as a function of sample limiting V-magnitude. V = 8 is taken as the completeness limit.

Table 1.

Adopted Stellar Masses, Main-Sequence Lifetimes, Luminosities,

Effective Temperatures, Bolometric Corrections, and Absolute Visual Magnitudes

| Mass ($M_\odot$) | $\tau_{MS}$ (Myr) | log L ($L_\odot$) | log $T_{eff}$ (K) | BC | $M_V$ |
|---|---|---|---|---|---|
| 85 | 2.823 | 6.006 | 4.705 | −4.52 | −5.77 |
| 60 | 3.447 | 5.728 | 4.683 | −4.37 | −5.23 |
| 40 | 4.303 | 5.373 | 4.640 | −4.08 | −4.63 |
| 25 | 6.408 | 4.897 | 4.579 | −3.66 | −3.86 |
| 20 | 8.141 | 4.650 | 4.544 | −3.42 | −3.48 |
| 15 | 11.584 | 4.303 | 4.492 | −3.07 | −2.97 |
| 12 | 16.018 | 4.013 | 4.448 | −2.76 | −2.55 |
| 9 | 26.389 | 3.617 | 4.383 | −2.32 | −2.00 |

Table 2.

Uncorrected (O3-B2) V Star Counts and Derived Results ($V_{lim} = 8$). See discussion of correction factors in sections 2.6 and 3.

| $M_V$ | $N(M_V)$ (stars) | $N(M_V, r_b)$ (stars) | $N_{eff}(M_V, r_b, \tau_f)$ (stars) | $\rho_{plane}$ (stars/kpc$^3$) | $r_{max}$ (kpc) | $\tau(M_V)$ (Myr) |
|---|---|---|---|---|---|---|
| −2.20 | 191.2 | 933.5 | 933.5 | 1447.7 | 0.683 | 21.974 |
| −2.60 | 39.3 | 155.5 | 226.1 | 241.1 | 0.760 | 15.110 |
| −3.00 | 88.7 | 283.9 | 565.2 | 440.2 | 0.844 | 11.036 |
| −3.35 | 26.8 | 71.2 | 179.9 | 110.4 | 0.926 | 8.697 |
| −3.56 | 0.3 | 0.6 | 1.7 | 0.9 | 0.979 | 7.638 |
| −3.70 | 27.5 | 60.8 | 189.8 | 94.3 | 1.016 | 7.039 |
| −3.90 | 7.6 | 15.1 | 52.8 | 23.5 | 1.071 | 6.300 |
| −4.10 | 11.9 | 21.3 | 82.7 | 33.1 | 1.129 | 5.674 |
| −4.25 | 2.0 | 3.3 | 13.8 | 5.1 | 1.175 | 5.265 |
| −4.40 | 5.2 | 7.9 | 35.4 | 12.2 | 1.222 | 4.900 |
| −4.60 | 2.0 | 2.7 | 13.5 | 4.3 | 1.289 | 4.472 |
| −4.80 | 7.0 | 8.6 | 46.4 | 13.4 | 1.359 | 4.098 |
| −4.90 | 5.2 | 6.1 | 33.9 | 9.4 | 1.395 | 3.930 |
| −5.00 | 2.3 | 2.6 | 15.1 | 4.0 | 1.432 | 3.771 |
| −5.05 | 1.0 | 1.1 | 6.4 | 1.7 | 1.451 | 3.696 |
| −5.10 | 1.0 | 1.1 | 6.4 | 1.6 | 1.471 | 3.623 |
| −5.20 | 2.0 | 2.0 | 12.6 | 3.1 | 1.510 | 3.483 |
| Totals | 421.0 | 1577.25 | 2415.12 | | | |

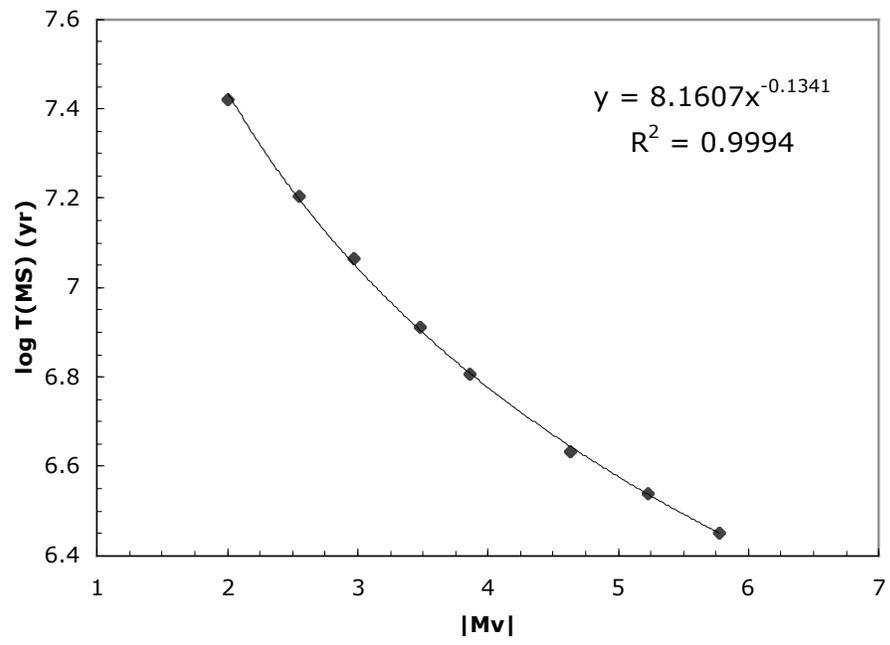

Figure 1

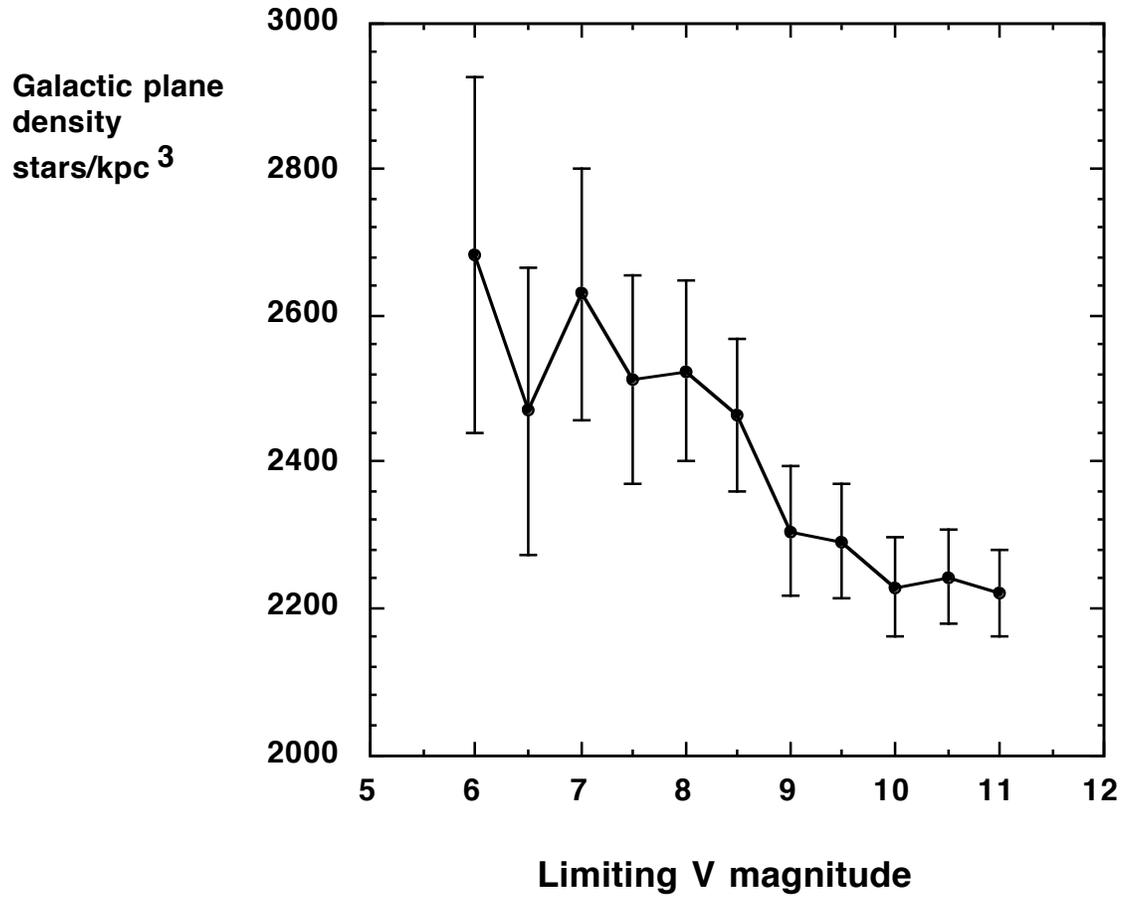

Figure 2